\documentclass[twocolumn,prb]{revtex4}%
\usepackage{amsmath}
\usepackage{amsfonts}
\usepackage{amssymb}
\usepackage{graphicx}
\usepackage{subfigure}
\usepackage[colorlinks,  linkcolor=blue, anchorcolor=blue,citecolor=blue,
urlcolor=blue ]{hyperref}
\usepackage{txfonts}%

\begin{document}

\title{Tuning of few-electron states and optical absorption anisotropy in GaAs quantum rings}   
\author{Zhenhua Wu$^{1\dag}$, Jian Li$^{2}$, Jun Li$^{3}$, Huaxiang Yin,$^{1}$ and Yu Liu$^{4\ddag}$ \footnote{$^{\ddag}$ liuyubj@inspur.com;\\$^{\dag}$wuzhenhua@ime.ac.cn.}}
\affiliation{$^{1}$Key Laboratory of Microelectronic Devices and Integrated
Technology, Institute of Microelectronics, Chinese Academy of Sciences,
100029 Beijing, China}
\affiliation{$^{2}$IQIS, College of Science, Chongqing University of Posts and
Telecommunications, Chongqing 400065, China}
\affiliation{$^{3}$Department
of Physics, Semiconductor Photonics Research Center, Xiamen University,
Xiamen 361005, China}
\affiliation{$^{4}$Inspur Electronic Information Industry Co., Ltd., Beijing 100085, China}

\begin{abstract}
The electronic and optical properties of a
GaAs quantum ring (QR) with few electrons in the presence of the Rashba
spin-orbit interaction (RSOI) and the Dresselhaus spin-orbit interaction
(DSOI) have been investigated theoretically. Configuration interaction (CI) method is employed to calculate the eigenvalues and eigenstates of the multiple-electron QR accurately. Our numerical results demonstrate that the symmetry breaking induced by the RSOI and DSOI
leads to an anisotropic distribution of multi-electron states. The Coulomb
interaction offers additional modulation of the electron distribution and
thus the optical absorption indices in the quantum rings. By tuning the
magnetic/electric fields and/or electron numbers in a quantum ring, one can
change its optical property significantly. Our theory provides a new way to
control multi-electron states and optical properties of QR by
hybrid modulations or by electrical means only.
\end{abstract}

\maketitle

\section{INTRODUCTION}

Recently all-electrical control of spin states have attracted extensive research interest in the quest for spintronics
and quantum information processing with high flexibility and controllability.\cite{Wolf,Tapash,Pekka,Ciorga,Areg}
The Rashba and Dresselhaus spin-orbit interactions\cite{Grundler,Nitta,Konig} (SOI)
provides us with an efficient way to control spin which has become one of
the most influential concepts in semiconductor spintronics since they are electrically tunable.~\cite{Zutic,Das} In the past few years, semiconductor quantum rings (QRs)
have attracted intensive interests due to their unique topological geometry
and energy spectrum,\cite{Junsaku,PeterFoldi,CDaday,AManaselyan,NKim} which opens new opportunities for realizing novel nano-scaled photonic detectors and sources, charge/spin memory cell, \emph{et.al.} Especially, The QRs behave like giant artificial atoms in a dielectric cage and show a multi-energy-level system in the presence of magnetic fields and/or SOI, thus presenting a potential source of qubit in quantum computing. Higher spin stability in QRs than in quantum dots, another candidate for qubit, make the relaxation and decoherence processes take place in the time scale that is sufficient long for spin manipulations.\cite{SBellucci,EZipper,VFomin} This is very attractive for the realization of spin qubits and throughout understanding of the multi-electron spin states in QRs is required.
State-of-the-art growth, etching and gate techniques have made it possible
to fabricate high-quality QRs, and control the number of electrons in a QR
exactly.\cite{AshooriRC,STarucha,CSomaschini,VTognetti} The ring-shaped charge distributions can also be realized in gate-all-around nanowire or core-shell nanowire arising from potential or geometrical confinement respectively.\cite{SH,GF,BW,AL,KP} The interplay between
the RSOI and DSOI results in a periodic potential in an isolated QR that
breaks the rotational symmetry, produces gaps in the energy spectrum and
suppresses the persistent currents.~\cite{SViefers,YangW1,YangW2,LCastelano} However, in these studies, one-electron models with effective mass approximation were used, without taking into account the many-body effects. In few-electron quantum confined system, the many-body effects and Coulomb interactions affect the electronic states significantly, which have attracted considerable interests.~\cite{AFuhrer,JClimente,YSaiga,LWagner,PLoos,CBall} So far the hybrid impacts of the RSOI, DSOI and the Coulomb interactions on the electronic and optical properties in a QR with few electrons have not been thoroughly investigated. Especially the competition between the spin-orbit interactions and the Coulomb interactions with increasing electron number in a QR from one to a relatively large number need to be addressed systematically.

\begin{figure}[tbp]
\scalebox{0.45}[0.45]{\includegraphics{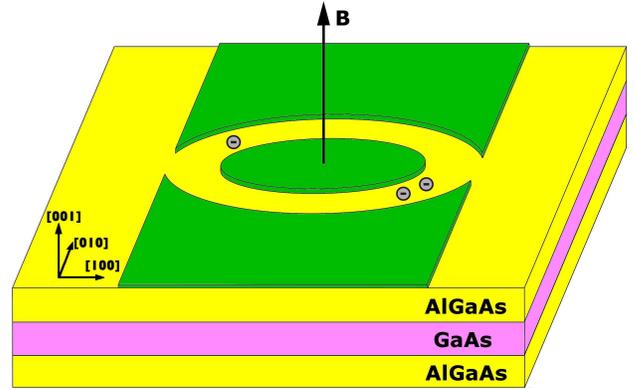}}
\caption{(Color online) Schematic diagram of a considered multi-electron QR in (001)-oriented GaAs/AlGaAs quantum well. The $x$ ($y$) axis is directed along the [100] ([010]) crystallographic orientation respectively. In this calculation, we consider an ideal one-dimensional QR ignoring the finite-width impacts and the coupling to the extended reservoirs.}
\label{Fig_1}
\end{figure}

In this study, we investigate theoretically the multi-electron states in
GaAs QRs in the presence of the SOIs and perpendicular magnetic/electric
fields.  Beyond the effective one electron models on similar systems, we employee the configuration interaction (CI) method, which is numerically
exact in this complex few electron system. The QR in the presence of the RSOI and DSOI behaves like two quantum
dots coupled laterally along specific crystallographic direction, i.e., [$110
$] or [$1\overline{1}0$]. The interdot coupling can be tuned by changing the
strengths of the SOIs. Interestingly, this anisotropic electron
distribution, which can be rotate from [$110$] to [$1\overline{1}0$]
direction by reversing the direction of the perpendicular electric field,
results in anisotropic optical properties that provide us with a possible
way to detect it experimentally.

\section{MODEL AND FORMULISM}

\subsection{SINGLE-ELECTRON HAMILTONIAN}
We consider a GaAs QR which can be fabricated in (001)-oriented symmetrical GaAs/AlGaAs quantum well as shown in Fig.~\ref{Fig_1}.
With coordinate axes directed along cubic crystallographic axes, i.e., $x || [100]$ direction and $y || [010]$ direction,
the single electron Hamiltonian with both RSOI and DSOI terms in a
finite-width QR under a perpendicular magnetic field is
given by,~\cite{FMeijer,Sheng,Liu,SGanichev}
\begin{eqnarray}
H &=&\frac{\hbar ^{2}k^{2}}{2m^{\ast }}+\alpha \left( \sigma
_{x}k_{y}-\sigma _{y}k_{x}\right) +\beta \left( \sigma _{x}k_{x}-\sigma
_{y}k_{y}\right) +\frac{1}{2}g^{\ast }\mu _{B}B\sigma _{z}  \notag \\
&&+V(r),
\end{eqnarray}
where $\vec{k}=-i\nabla +e\vec{A}/\hbar $. $\vec{A}\left( \vec{r}%
\right) =B/2(-y,x,0)$ is the vector potential. $m^{\ast }$ is the electron
effective mass. The fourth term describes the Zeeman splitting with Bohr
magneton $\mu _{B}=e\hbar /2m_{0}$ and the effective $g$ factor $g^{\ast }$.
$\sigma _{x}\left( i=x,y,z\right) $ are the Pauli matrices. $\alpha (\beta )$
specify the RSOI (DSOI) strengths. To include the RSOI and DSOI, an additional up-down asymmetry need to be present by, e.g., electric field applied normally to QR plane. Consequently the rotation symmetry of the system is reduced to $C_{2v}$. $V(r)$ is the radial confining potential.
We adopt hard-wall boundary conditions which can be guaranteed by the strong confinement potential,
\begin{equation}
V(r)=\left\{
\begin{array}{c}
0, \text{\ \ \ \ \ \ \ \ \ r = R}\ \ \ \\
\infty ,\text{ \ \ \ otherwise}\ \ \
\end{array}%
\right.
\end{equation}%
where $R$ is radii of the quantum ring. In the presence of both RSOI and DSOI, the single-electron dimensionless 1D Hamiltonian is written as, \cite{Sheng,Liu}
\begin{eqnarray}
H_{e} &=&\left[ -i\frac{\partial }{\partial \varphi }+\frac{\Phi }{\Phi _{0}}%
+\frac{\overline{\alpha }}{2}\sigma_{r}(r,\varphi)-\frac{\overline{\beta }}{2}\sigma_{\varphi}(r,- \varphi )\right] ^{2}\notag \\
&&-\frac{\overline{\alpha }^{2}+\overline{%
\beta }^{2}}{4}
+\frac{\overline{\alpha }\overline{\beta }}{2}\sin 2\varphi +\frac{1}{2}%
\overline{g}b\sigma _{z},
\label{He}
\end{eqnarray}%
where $\sigma_{r}(r,\varphi)=\cos \varphi \sigma _{x}+\sin \varphi \sigma _{y}$, $%
\sigma_{\varphi}(r,- \varphi )=\cos \varphi \sigma _{y}+\sin \varphi \sigma _{x}$,
where $\sigma_{x}$, $\sigma_{y} $ are Pauli matrices, $\Phi
=B\pi R^{2}$ is the magnetic flux threading the ring, $\Phi _{0}=h/e$ is the
flux unit, $b=e\hbar B/m^{\ast }E_{0}= 4 \Phi/ \Phi_0 $ is the dimensionless magnetic field,
$\overline{\alpha }(\overline{\beta })=\alpha (\beta )/E_{0}R$ specifies the
dimensionless RSOI (DSOI) strength, $E_{0}=\hbar ^{2}/2m^{\ast }R^{2}$ with
the ring radius $R$, and $\overline{g}=g^{\ast }m^{\ast }/2m_{0}$ is the
dimensionless $g$ factor.

\subsection{FEW-ELECTRON HAMILTONIAN}
The total Hamiltonian of the multi-electron QR can be rewritten in
second-quantization
\begin{eqnarray}
\mathcal{H} &=&\sum\limits_{i }E_{i }a_{i }^{+}a_{i }  \notag  \label{Hamiltonan} \\
&&+\frac{1}{2}\gamma \sum\limits_{i j i ^{^{\prime }}j ^{^{\prime }}}\langle
i j |U|i ^{^{\prime }}j ^{^{\prime }}\rangle a_{i }^{+}a_{j }^{+}a_{i
^{^{\prime }}}a_{j ^{^{\prime }}},
\end{eqnarray}%
where $|ij\rangle =a_{i}^{+}a_{j}^{+}|0\rangle $, $a_{i}^{+}  (a_{i})$ is the electron creation (annihilation) operator of the
states. $i$ and $j$ denoting the ith and jth single-electron energy
states, respectively, $E_{i }$~is the~energy of the $i $-th single electron level,which can
be obtained numerically by solving the single electron Schr\"{o}dinger
equation shown in Eq.~\ref{He}. The parameter~$\gamma =e^{2}/4\pi \varepsilon
_{0}\varepsilon $ and~$U=1/r$, where $r$ is the electron-electron distance.

The full configuration interaction (CI) method has been widely used to solve the many-body, non-relativistic Schr\"{o}dinger equation~\cite{JForesman,DCasanova,MNeff,DAlcoba,JKim,JSchriber} rather than the ab-initio approaches~\cite{Ishikawa}.
We adopt the CI method with employing adequate single-particle electron states, to calculate the eigenvalues and eigenstates of the above
Hamiltonian. The total wavefunction can be expanded $|\chi \rangle
=\sum\limits_{I}C_{I}|I\rangle $. The two electrons configuration is~$%
|I\rangle =|\cdots 01_{i}0\cdots 01_{j}0\cdots \rangle
=a_{i}^{+}a_{j}^{+}|0\rangle $ with $i<j$, (or ~$%
|J\rangle =|\cdots 01_{p}0\cdots 01_{q}0\cdots \rangle
=a_{p}^{+}a_{q}^{+}|0\rangle $ with $p<q$.). Here $|0\rangle $~represents the
vacuum state and~$a_{i}^{+}$ $(a_{i})$~is the electron creation
(annihilation) operator of the states. $i$ and $j$~denoting the $i$-th and~$j
$-th single electron energy states, respectively. The matrix element of the
total Hamiltonian for two electrons can be calculated
\begin{equation}
\langle I|\mathcal{H}|J\rangle =(E_{i}+E_{j})\delta _{ip}\delta _{jq}+\gamma \lbrack
\langle ij|U|qp\rangle -\langle ij|U|pq\rangle ].
\end{equation}
The N-electron configuration is $\left\vert M\right\rangle =\left(
a_{1}^{\dagger}\right) ^{n_{1}}\left( a_{2}^{\dagger}\right)
^{n_{2}}\cdots\left\vert 0\right\rangle $, where $n_{1},n_{2},\cdots$ are
either 1 or 0, and satisfied $\sum\limits_{i=1}^{\infty}n_{i}=N$. Then the
matrix element of the total Hamiltonian for $N$ electrons is given by,
\begin{align}
\left\langle M\left\vert \mathcal{H} \right\vert M^{\prime}\right\rangle & =\left( E_{%
\overline{n}_{1}}+E_{\overline{n}_{2}}+E_{\overline{n}_{3}}+\cdots +E_{%
\overline{n}_{N}}\right) \delta_{\left\langle n\right\rangle \left\langle
n^{\prime}\right\rangle }  \notag \\
& +\sum\limits_{q=1}^{Q}\left\{
\begin{array}{c}
\gamma \left\langle m_{dq}m_{cq}\left\vert U\right\vert m_{aq}m_{bq}\right\rangle
\\
-\gamma \left\langle m_{cq}m_{dq}\left\vert U\right\vert m_{aq}m_{bq}\right\rangle%
\end{array}
\right\}  \notag \\
& \times\left( -1\right) ^{n_{db,q}+n_{ca,q}},
\end{align}
where $\overline{n}_{i}$ identify the position of nonzero $n_{i}$, $%
\delta_{\left\langle n\right\rangle \left\langle n^{\prime}\right\rangle
}=\prod\limits_{i}\delta_{\overline{n}_{i},\overline{n^{\prime}}%
_{i}}=\delta_{MM^{\prime}}$. The factor $\left( -1\right)
^{b_{q}-a_{q}+d_{q}-c_{q}}\equiv\left( -1\right)
^{b_{q}-a_{q}-d_{q}+c_{q}}\equiv\left( -1\right) ^{n_{db,q}+n_{ca,q}}$ comes
from moving two electrons from states $m_{aq}$, $m_{bq}$ to $m_{cq}$, $m_{dq}$%
. $n_{db,q}=d_{q}-b_{q}+1$ ($n_{ca,q}=c_{q}-a_{q}+1$) is the number of
electrons in the states $m_{c}$ and $m_{a}$ ($m_{d}$ and $m_{b}$) respectively.  The
number $Q$ obeys the following rules
: (1) If there are four $m_{i}$ differing
between  $\left\vert M\right\rangle$
and $\left\vert M^{\prime} \right\rangle$, there is only one choice, $Q=1$. (2) If
only two $m_{i}$ differing
between  $\left\vert M\right\rangle$
and $\left\vert M^{\prime} \right\rangle$, $Q=N-1$. (3) If $\left\vert M\right\rangle = \left\vert M^{\prime} \right\rangle$, $%
Q=N(N-1)/2$. Solving the above secular equation, we can obtain the
eigenenergies and the eigenstates of the multi-electron system, and
thus calculate the electron distributions, and optical property of the
multi-electron QRs.

\subsection{OPTICAL ABSORPTION}

The optical absorption rate is obtained within the electric-dipole
approximation%
\begin{eqnarray}
W &=&\frac{2\pi }{\hbar }{\sum\limits_{f}}|\langle f|H_{ep}|i\rangle
|^{2}\delta (E_{f}-E_{i}) \\
&=&\frac{2\pi }{\hbar }\sum\limits_{f,i}(\frac{eA_{0}}{m^{\ast }}%
)^{2}|\langle f|\overset{\rightharpoonup }{\epsilon _{\lambda }}\cdot (%
\overset{\rightharpoonup }{p_{1}}+\overset{\rightharpoonup }{p_{2}}%
)|i\rangle |^{2}  \notag \\
&&\times \delta (E_{f}-E_{i}-\hbar \omega ),  \notag
\end{eqnarray}%
where $E_{i}$~and~$E_{f}$~are the energies of the initial and final states,
respectively. $H_{ep}=(e/m^{\ast })(\overset{\rightharpoonup }{A}\cdot
\overset{\rightharpoonup }{p_{1}}+\overset{\rightharpoonup }{A}\cdot \overset%
{\rightharpoonup }{p_{2}})$, here ~$\overset{\rightharpoonup }{p}_{1,2}$~are
the canonical momenta of the electrons and $\overset{\rightharpoonup}{A} =%
\displaystyle\sum\limits_{k} A_{0}\overset{\rightharpoonup}{%
\epsilon_{\lambda}}\{a_{k\lambda}e^{i(-\omega t+\overset{\rightharpoonup}{%
k_{\lambda}}\cdot\overset{\rightharpoonup}{r} )}+a_{k\lambda}^{+}e^{i(\omega
t-\overset{\rightharpoonup}{k_{\lambda}} \cdot\overset{\rightharpoonup}{r}%
)}\}$ in which $a_{k\lambda}$, $a_{k\lambda}^{+}$ are photon annihilation
and creation operators, respectively.  $A_{0}$,~$\omega $~and~$\overset{%
\rightharpoonup}{\epsilon }_{\lambda } $~are the amplitude, frequency and
polarization vector of the incident linear-polarized light. In the
calculation we replace the energy delta function $\delta (E_{12}-\hbar
\omega )$ with a Lorentz broadened function $(\Gamma /\pi )/((\hbar \omega
-E_{12})^{2}+\Gamma ^{2})$ where~$\Gamma$ is the broadening parameter
describing the homogeneous broadening of the energy levels in the ring.
Then we can obtain the experimentally measurable absorption index $A$, which is defined by the following light damping equation,
\begin{equation}
F(x)=F_{0}e^{-A x},
\end{equation}
where $F_{0}$ is the flux at $x=0$. The energy flux equation is
$\nabla\cdot\vec{F}+\frac{\partial n_{\lambda}}{\partial t}=0$
, and the flux can also be written as $F=\upsilon_{g}n_{\lambda}$ , here
$\upsilon_{g}=c/\eta$ is the speed of energy flux and $\eta$ is the refraction index
of the medium. Thus the absorption index $A$ can be written as,
\begin{eqnarray}
A &=& \frac{1}{F}\frac{dn_{\lambda}}{dt}=\frac{W}{\upsilon_{g}n_{\lambda}} \\
&=& \frac{2\pi}{\hslash}\left(  \frac{eA_{0}}{m^{\ast}}\right)  ^{2}%
\sum_{f,i}\frac{\left\vert \left\langle f\left\vert \vec{\epsilon}_{\lambda
}\cdot\left(  \vec{p}_{1}+\vec{p}_{2}\right)  \right\vert i\right\rangle
\right\vert ^{2}}{c/\eta}\delta\left(  E_{fi}-\hslash\omega\right) \notag
\end{eqnarray}

\section{RESULTS AND DISCUSSIONS}

The accuracy of the calculated multi-electron energy spectrum depends on the
number of possible many-particle configurations that are used which is
determined by the number of electrons ($N_{e}=3, 4, 5,$~and~$6$~in this
paper), and the number of single-particle states $N_{S}$. In the calculation
we include~$30$~single-particle electron states to ensure that the lower
multi-electron states are numerically accurate, for instance, the accuracy
of the lowest levels can approach to $1.0 \times 10^{-4}meV$. For
simplicity, all physical quantities are taken dimensionless, e.g., the
length unit is the radius of the ring~$R$, the energy unit is $E_{0}$ and
the magnetic field unit is $b$. The relevant parameters for GaAs are:~\cite%
{Madelung} the electron effective mass~$m^{\ast }=0.067m_{0}$, and the dielectric constant~$\varepsilon =12.5$. For an example, for $R=30$ nm we find $E_0=0.633$ meV and $b=2.73$ when $B=1$ T. 
In this study, we focus on the interplay between RSOI, DSOI and the Coulomb interactions with ignoring the Zeeman term.

\begin{figure}[tbh]
\centering
\includegraphics[width=0.9\columnwidth]{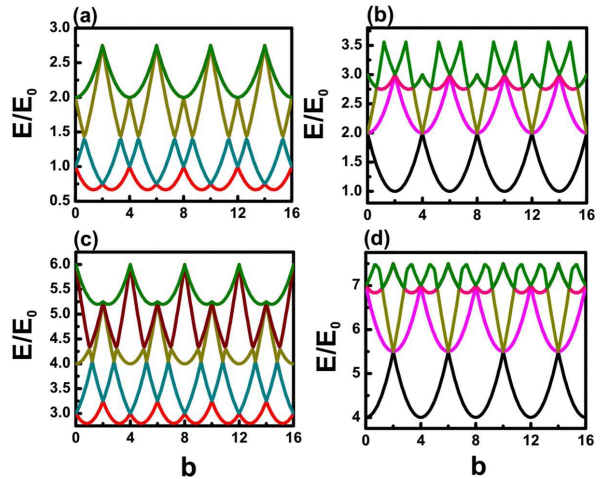}
\caption{(Color online)The energy spectrum for 1D GaAs rings without
Coulomb interaction and SOIs for (a) $N_{e}$ = 3, (b) $N_{e}$ = 4, (c) $N_{e}$ = 5, and (d) $N_{e}$ = 6,
 respectively.}
\label{E_B1}
\end{figure}

First we consider three electrons in a 1D GaAs ring. The electron states are obtained by coupling the well known singlet and triplet two-electron states with a third electron. There are total of $2^{3}=8$ spin states including one set of quartet states and two sets of doublet states that involve two pairs of
total spin 1/2 states built up of a singlet and an unpaired spin. The symmetries of both the spatial
wavefunctions and the spin states which finally make an antisymmetric
wavefunction as requested by the Pauli principle.
In Fig.~\ref{E_B1} we plot the low energy dispersion relations of a 1D GaAs ring with
$N_{e}=3,4,5,6$ electrons without considering the SOIs or the Coulomb interactions. For a QR with three electrons but without Coulomb interactions and SOIs, we plot the lowest few energy levels versus a perpendicular magnetic field in Fig.~\ref{E_B1}(a). One can find that the ground energy level lies in the degenerated doublet states in the absence of an external perpendicular magnetic field. The first exited energy level lies in the quartet states. As the magnetic field increases, the degeneracy is lifted due to the effect of the magnetic field on Landau level. The energy spectra exhibit a periodic dependence on the dimensionless magnetic field $b$ accounting for the moving of Landau levels. For a QR including more than three electrons, similar low-energy spectra are obtained following the same considerations. Interestingly we find that the parabolic and crossing or anticrossing features in energy levels of are similar for set of (1, 3, 5) electrons and the other set of (2, 4, 6) electrons alternatively (see Fig.~\ref{E_B1} and Refs.~\cite{Sheng,Liu} for a QR with 1 or 2 electrons). The two sets of multi-electron QRs arising from the alternatively changed basis configurations given in Table~\ref{eleconfig}.

\begin{table}%
\begin{tabular*}
{0.9\columnwidth}[c]{@{\extracolsep{\fill} }ccccc}
\hline\hline
For two electrons& $S=0$    & $S=1$   &   &
 \\
 &  Singlet &  Triplet &   &
 \\
\hline
For three electrons & $S=1/2$    & $S=3/2$   &   &
\\
  & Doublet & Quartet &   &
\\
\hline
For four electrons & $S=0$    & $S=1$   & $S=2$  &
\\
  & Singlet &  Triplet & Quintet  &
\\
\hline
For five electrons & $S=1/2$    & $S=3/2$   & $S=5/2$  & etc.
\\
  & Doublet & Quartet  &Sextet  &
\\
\hline
For six electrons & $S=0$    & $S=1$   & $S=2$  & etc.
\\
  & Singlet &  Triplet & Quintet  &
\\
\hline  \hline
\end{tabular*}
\caption{Basis electronic configurations of 2 to 6 identical electrons in a QR.}%
\label{eleconfig}%
\end{table}

\begin{figure}[tbh]
\centering
\includegraphics[width=0.95\columnwidth]{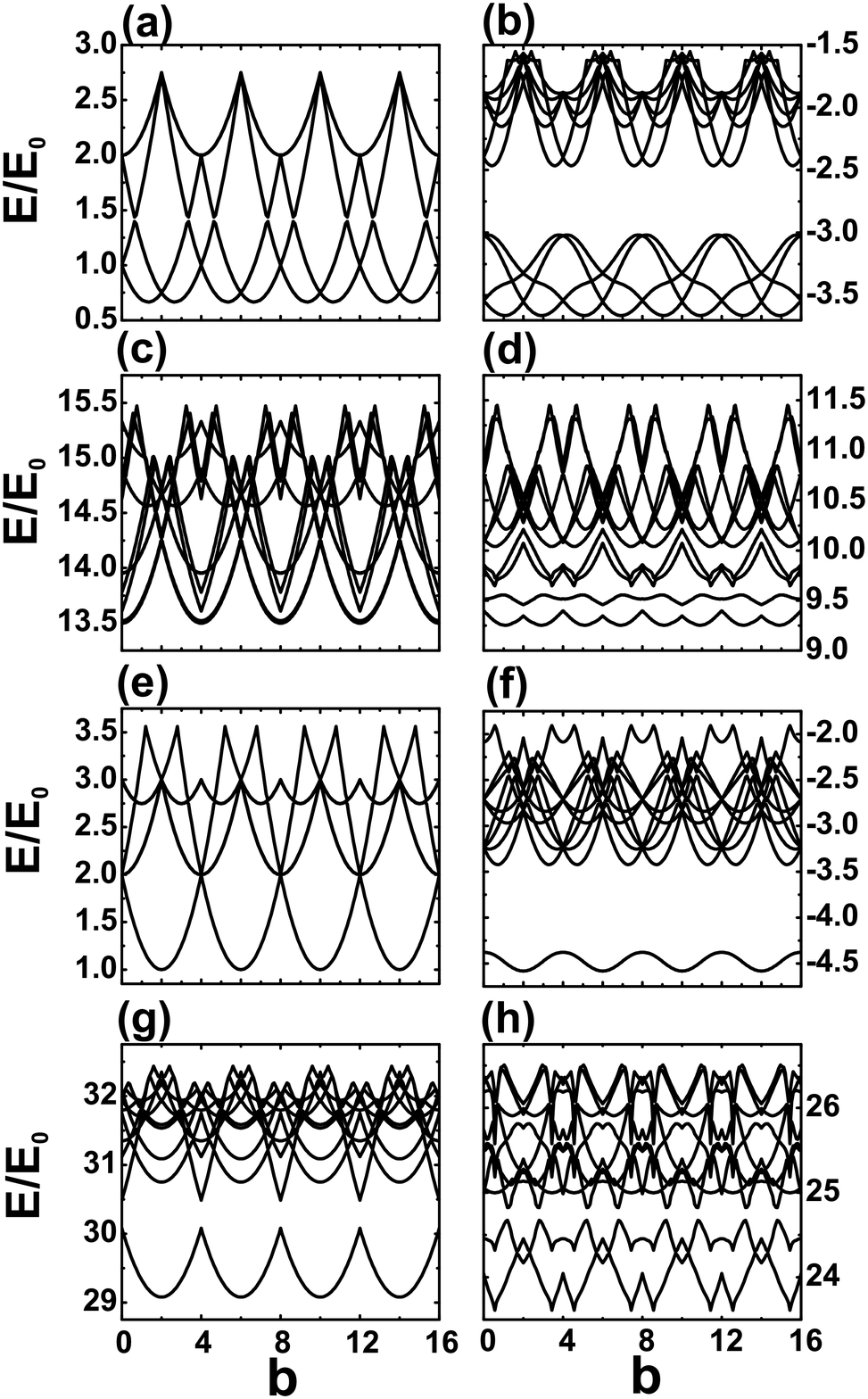}
\caption{(Color online) The energy spectrum for 1D GaAs rings, (a) $N_{e}=$ 3 electrons
without Coulomb interaction and SOIs, (b) $N_{e}=$ 3 electrons without Coulomb interaction but with SOIs, (c) $N_{e}=$ 3 electrons with Coulomb interaction but without SOIs, (d) and $N_{e}=$ 3 electrons with both
Coulomb interaction and SOIs. (e)-(h) The sames with (a)-(d) but with $N_{e}=$ 4 electrons. ~$ \overline{\protect\alpha}=2.0, \overline{\protect\beta}=1.0$.}
\label{E_B}
\end{figure}

The SOIs and Coulomb interactions play important roles in the multi-electron system. We also start by exploring the energy spectra with three electrons in a 1D GaAs ring.
When the SOIs are induced, the eigenvalues of spin operator $\hat{S}^{2}$ are no longer good quantum numbers, the singlet and triplet states are mixed due to spin-orbit coupling. The energy levels of the three-electron QR get lower, and the spin degenerate is lifted as shown in Fig.~\ref{E_B}(b).
Since the SOIs reduce the total energy of the three-electron system as determined by the Hamiltonian. The RSOI and DSOI with different strengths in our calculations, i.e.,~$\overline{\alpha}=2.0, \overline{\beta}=1.0$, break the
symmetry of the 1D Hamiltonian. Then the spin-up and spin-down electron energy levels with the same quantum number are separated.
For comparison, the Coulomb interactions are considered without inducing SOIs, the energy levels of the three-electron QR are increased significantly,
since the Coulomb interactions increase the repulsive energy. Importantly the Coulomb exchange interactions give rise to the
splitting of the triplet states $S_{z}=0$ and $S_{z}=\pm1$ and the coupling of the doublet and quartet states. We therefore observe complex crossings and anticrossings in the energy spectrum as shown in Fig.~\ref{E_B}(c).
The interplay of both Coulomb interaction and the SOIs destroy the parabolic energy dispersions and lift the spin degeneracy in the three-electron QR due to the competition between the two kinds of interactions, resulting in very complicated energy spectra as shown in Fig.~\ref{E_B} (d). For a QR with more than
three electrons, above considerations are still valid. Accounting for the configuration of singlet, triplet and quintet states instead of doublet and quartet states, the feature of energy spectrum is modulated by adding the fourth electron in the QR as shown in
Fig.~\ref{E_B} (e-h). The oscillations length $b$ is doubled,
this can be readily explained in terms of the occupation of an even or odd number of single-electron levels.

\begin{figure}[tbh]
\centering
\includegraphics[width=0.95\columnwidth]{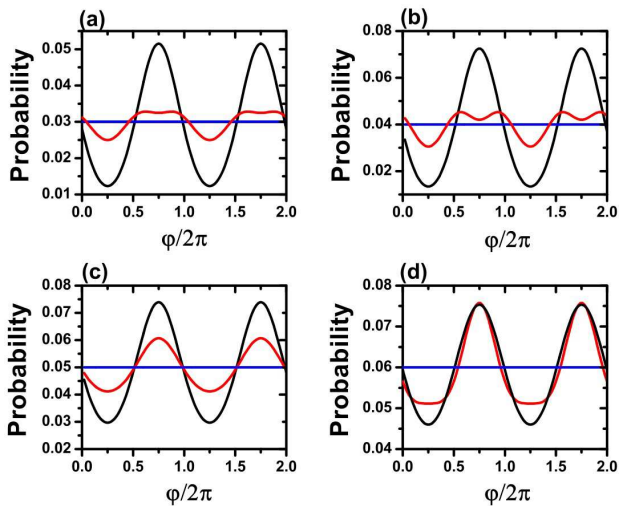} %
\caption{( The ground state distribution in a ring as a function of angular~$%
\protect\varphi$ with (a) 3 electrons, (b) 4 electrons (c) 5 electrons (d) 6
electrons. Blue lines: with Coulomb interaction but without SOIs. Black
lines: without Coulomb interactions but with SOIs. Red lines: with both
Coulomb interactions and SOIs. }
\label{1D}
\end{figure}

Next we investigate the electron distributions in the GaAs QR and the dominating physical mechanism. Fig.~\ref{1D}(a)-(d) show the exclusive or hybrid impacts of Coulomb interaction and SOIs on the electron distributions in the QR containing three, four, five and six electrons respectively. The Coulomb repulsion makes
the multiple electrons tend to avoid each other and localize equally separated in the
ring but at any crystallographic direction as indicated by blue lines in Fig.~\ref{1D}.
The interplay between the RSOI and DSOI breaks the
rotational symmetry and results in an azimuthal periodic potential, whose height is determined by the product of the strengths of RSOI and DSOI in Eq.~\ref{He}. From this Hamiltonian, one can see
clearly that the QR with RSOI and DSOI does not possess the rotational
symmetry, because there are two potential wells at $\varphi =3\pi /4$ and $\varphi
=-\pi /4$.
Consequently, it leads to an azimuthal anisotropic
electron distribution. The electron distribution of the ground
state shows a bar-bell-like shape along the specific crystallographic
directions at $\varphi =3\pi /4$ and $\varphi
=-\pi /4$. When both Coulomb interaction and SOIs are incorporated in Eq.~\ref{He}, the competition between the two type interactions gives rise to complicate electron
distribution which differs when adding more electrons and/or applying a finite perpendicular magnetic field. The QR behaves like a laterally-coupled double quantum dots.
In the following we illustrate in more details the impact of magnetic fields on the electron distributions in the QR.

\begin{figure}[tbh]
\includegraphics[width=1.0\columnwidth]{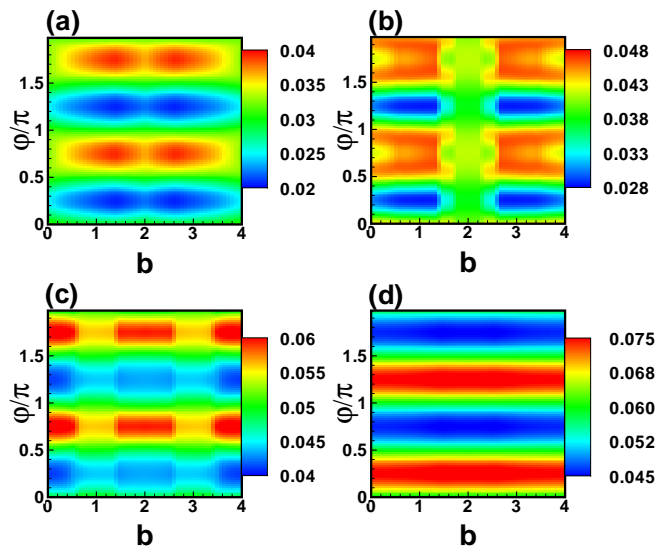}
\caption{(Color online) Contour plot of the electron distribution in (a)
three-electron QR, (b) four-electron QR, (c) five-electron QR, (d)
six-electron QR as function of magnetic field~$b$ and the crystallographic
direction~$\protect\phi$. ~$\overline{\protect\alpha}=2.0, \overline{\protect%
\beta}=1.0$.}
\label{EleDistr}
\end{figure}

Figure~\ref{EleDistr} depicts the contour plot of the electron
distribution in a multi-electron QR of different cases in the presence of both
the SOIs and the Coulomb interaction. One can see that the electrons are
mainly localized in a region along the specific crystallographic direction~$%
[1\overline{1}0]$~which makes the single QR behave like laterally-coupled
double quantum dots. This is caused by the interplay between the RSOI and
DSOI which breaks the rotational symmetry and results in an azimuthal
periodic potential.~\cite{Sheng} The behavior of electron distribution is
very different with increasing magnetic field~$b$~for different electron
numbers. This is caused by the differences of overlap between electrons of
each single electron distribution. For three-electron case there are two
kinds of possible single electron distribution: (i) two electrons are
localized at $\varphi=3\pi/4$ or $\varphi=-\pi/4$, while the third electron
localizes at the opposite side, i.e., $\varphi=-\pi/4$ or $\varphi=3\pi/4$; (ii)
all three electrons localize at $\varphi=3\pi/4$ or $\varphi=-\pi/4$. Due to the
repulsive Coulomb interaction, this configuration has higher energy,
therefore the ground state of three-electron case prefers a triangular
configuration. This is because the azimuthal confining potential (the third
term in Hamiltonian (1)) squeezes the electron wavefunctions and forces them
align along the specific crystallographic direction, i.e., $[1\overline{1}0]$
(see Fig.~\ref{EleDistr}(a)), while the repulsive Coulomb interaction pushes
two electrons away from this direction. Therefore one can see clearly that
the peaks of electron distribution at $\varphi=3\pi/4$ or $\varphi=-\pi/4$ become
broadening comparing with the two-electron case (see Fig. 7 in Ref. [%
\cite{Liu}]). For four and five electron QR they both have three kinds
of possible single electron distribution which give more complicated
features shown in Figs.~\ref{EleDistr}(b) and \ref{EleDistr}(c). However,
the six-electron QR shows in Fig.~\ref{EleDistr}(d) releases a feature very
similar with the two-electron's.~\cite{Liu} This is because the repulsive
Coulomb interaction makes the ground state of six-electron case prefer a
bar-bell like configuration, i.e., three electrons are localized at $%
\varphi=3\pi/4$ or $\varphi=-\pi/4$, while the other three localize at the
opposite side. In this case the three electrons are too close and therefore
the density distribution (see Fig.~\ref{EleDistr}(d)) shows a single peak
but with width broadening (see the color scales of Fig.~\ref{EleDistr}).We can switch the potential minima from $[110]$ to
$[1\overline{1}0]$ easily by reversing the direction of the perpendicular
electric field, i.e., $\overline{\alpha}$ to $-\overline{\alpha}$. The
orientation of the electron distribution can be switched from~$[1\overline{1}%
0]$~to~$[110]$. Besides, by tuning the external magnetic field~$b$, we can
tune the shape of the electron distribution. Thus, it provides a method to
control the electron state by using SOIs and the external magnetic field.

\begin{figure}[tbh]
\includegraphics[width=1.0\columnwidth]{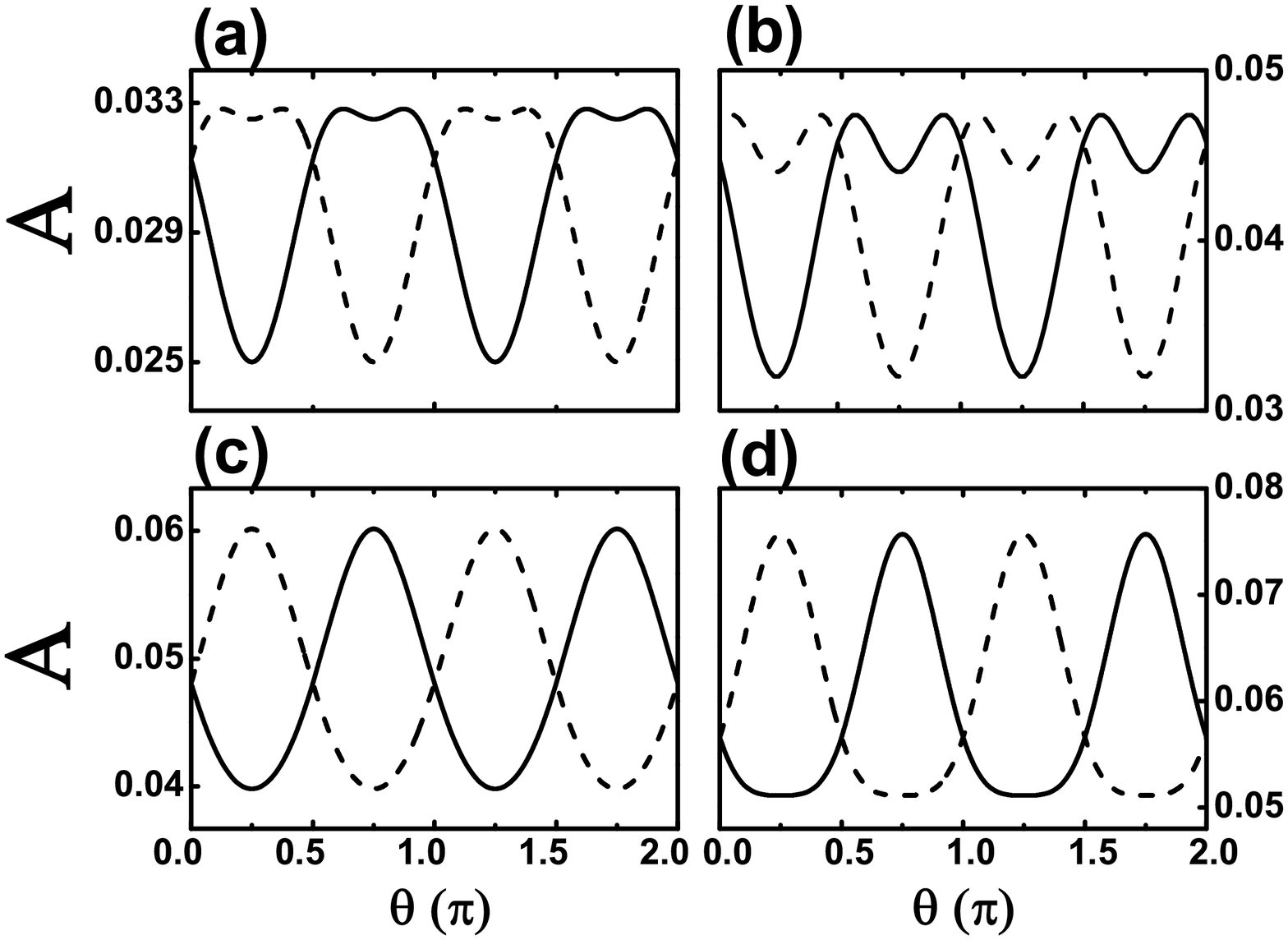} %
\caption{The optical absorption indices $a$ as a function of the direction
of the polarized vector of the incident linear polarized light of (a)
three-electron QR, (b) four-electron QR, (c) five-electron QR, (d)
six-electron QR.~$\overline{\protect\beta}=1.0$, $\overline{\protect\alpha}%
=2.0$ (the solid line), $\overline{\protect\alpha}=-2.0$ (the dashed line). The broadening parameter is fixed $\Gamma = 1/\pi^{2}$.}
\label{Optical}
\end{figure}

However the electron states and electron distributions are difficult to measure directly.
We therefore propose an optical measurement method to detect the multi-electron spatial
anisotropic distribution by monitoring the optical absorption in the infrared regime. The anisotropic electron distribution and the corresponding overlap factor between the ground and first excited states gives rise to anisotropic absorption. We consider a beam of
linear-polarized light incident along the $z$ axis and calculate the
optical absorption in the infrared regime. In Fig.~\ref{Optical} we
plot the optical absorption for different electron numbers as a function of
the angle $\theta $ of the polarization plane of the incident
linear-polarized light with respect to the $x$ axis at zero magnetic field.
The optical absorption oscillates periodically which is
consistent with the electron distributions accounting for their dependence on crystallographic angle $\theta $ that shown in Fig.~\ref{EleDistr}%
. As we observed before, the RSOI and DSOI and the Coulomb repulsion between these electrons make the localized electron distribution in the ring along the crystallographic direction $\varphi =\pm \pi /4$ with different shape. These anisotropic
electron distribution is in good agreement with the anisotropic behavior of the
optical absorption, i.e., related oscillating dependence of the optical absorption
an the crystallographic direction $\theta$ as shown in Fig.~\ref%
{Optical}). By reversing the direction of the perpendicular electric field,
i.e., $\overline{\alpha}$ to $-\overline{\alpha}$, the optical absorption
will change strongly since the electric dipoles of these electrons change
from parallel to perpendicular with respect to the polarization vector of
the incident light. This distinct variation provides us an efficient tool
to detect the anisotropy in the electron distribution.

\section{CONCLUSIONS}
In summary, we have studied multi-electron energy spectra in a GaAs QR by using the CI method. In our calculation, the Coulomb interaction, Rashba SOI and Dresselhaus SOI are explicitly included. We demonstrate theoretically the anisotropic distribution
of multi-electron states in a semiconductor quantum ring. Our numerical results illustrate that the energy spectra and the charge distributions are controlled by
the interplay between the Rashba SOI, Dresselhaus SOI and the Coulomb
interaction in the presence of perpendicular magnetic/electric fields. We propose a possible experimental manifestation that the anisotropy in multi-electron QR can be switched by reversing the direction of the perpendicular
electric field and be detected by the optical absorption measurement.

\begin{acknowledgments}
This work was supported by the MOST (Grants No. 2016YFA0202300), the NSFC (Grants No. 11604036) and the Opening Project of MEDIT, CAS. Jun Li was supported by the Natural Science Foundation of Fujian Province of China (Grant No. 2016J05163) and the Fundamental Research Funds for the Central Universities (Grant No. 20720160019).

\end{acknowledgments}

\end{document}